\documentclass{article}
\usepackage{times}
\usepackage{amsfonts}
\usepackage[pdfmark,colorlinks]{hyperref}
\begin{document}
\title{Bypassing Pauli's Theorem}
\author{Jos\'e M. Isidro\\
Instituto de F\'{\i}sica Corpuscular (CSIC--UVEG)\\
Apartado de Correos 22085, Valencia 46071, Spain\\
{\tt jmisidro@ific.uv.es}}

\maketitle

\begin{abstract}

We define a quantum--mechanical time operator that is selfadjoint and compatible with the energy operator having a spectrum bounded from below. On their common domain, the operators of time and energy satisfy the expected canonical commutation relation. Pauli's theorem is bypassed because the correspondence between time and energy is not given by the standard Fourier transformation, but by a variant thereof known as the holomorphic Fourier transformation.

\end{abstract}

\tableofcontents

\section{Introduction}\label{summary}

The definition of a time operator in quantum mechanics is an outstanding problem ever since Pauli's theorem \cite{PAULI}; see ref. \cite{HOLLAND} 
for a brief account and ref. \cite{TIME} for a detailed treatment. This has prompted attempts to introduce arrival time and time--of--flight operators \cite{EGUSQUIZA}, and to provide a physical interpretation for quantum theories based on nonhermitian operators \cite{QUASI} or on positive, operator--valued measures \cite{POVM}. Critical assessments of the technical aspects of Pauli's theorem have also appeared \cite{GALAPON}.

In this letter we present an alternative definition of a quantum--mechanical time operator that bypasses the technical objections raised by Pauli to the existence of a quantum--mechanical time operator. Briefly, a selfadjoint Hamiltonian operator $H$ that is bounded from below is placed in canonical correspondence with a nonhermitian time operator $T$ via the {\it holomorphic Fourier transformation}\/ (HFT) \cite{RUDIN}. The latter differs substantially from the standard Fourier transformation used in quantum mechanics. Perhaps its most striking feature is the appearance of a nonhermitian time operator $T$ that is canonically conjugate, via the HFT, to the Hamiltonian $H$. However, the square $T^2$ admits a selfadjoint Friedrichs extension $T^2_F$. Finally, $T^2_F$ admits a selfadjoint square root that serves as a {\it bona fide}\/ time operator. After the technical presentation of sections \ref{hft} and \ref{qmhft}, we present some examples and discuss our conclusions in section \ref{discussion}.

\section{The holomorphic Fourier transformation}\label{hft}

Background material on the HFT, summarised in the following, can be found in ref. \cite{RUDIN}.
Let  $\mathbb{H}$ denote the upper half plane: the set of all $z\in\mathbb{C}$ such that Im$(z)>0$. Let $f\in L^2(0,\infty)$. For
$z=x+{\rm i} \,y\in \mathbb{H}$, the function $\varphi$ defined as
\begin{equation}
\varphi(z):={1\over\sqrt {2\pi}}\int_0^{\infty}{\rm d} s \,f(s) \, {\rm e}^{{\rm i} sz},
\label{hfourier}
\end{equation}
the integral understood in the sense of Lebesgue, is holomorphic on $\mathbb{H}$. 
Its restrictions to horizontal straight lines $y={\rm const}>0$ in $\mathbb{H}$ 
are a bounded set in $L^2(\mathbb{R})$.

Conversely, let $\varphi$ be holomorphic on $\mathbb{H}$, and assume that 
\begin{equation}
{\rm sup}_{ 0<y<\infty}\, \int_{-\infty}^{\infty} {\rm d} x \,|\varphi(x+{\rm i} y)|^2=C<\infty.
\label{sup}
\end{equation}
Then the function $f$ defined by      
\begin{equation} 
f(s):={1\over \sqrt {2\pi}}\int_{-\infty}^{\infty}{\rm d} z \,\varphi (z)\, {\rm e}^{-{\rm i}sz},
\label{ihfourier}
\end{equation}
the integration being along any horizontal straight line $y= {\rm const}>0$ in $\mathbb{H}$,
satisfies the following properties. First, $f(s)$ is independent of the particular
horizontal line $y= {\rm const}>0$ chosen. Second, $f\in L^2(0,\infty)$. Third, 
for any $z\in \mathbb{H}$, eqn. (\ref{hfourier}) holds, with 
\begin{equation}
\int_{0}^{\infty}{\rm d} s \,|f(s)|^2 = C.
\label{norma}
\end{equation}
We call $f$ the holomorphic Fourier transform of $\varphi$.

Some features of the HFT on $\mathbb{H}$ are worth mentioning. Let $\Omega(\mathbb{H})$ 
denote the space of all holomorphic functions on
$\mathbb{H}$, and let $\Omega_0(\mathbb{H})$ denote the proper subspace of all
$\varphi\in\Omega(\mathbb{H})$ such that the supremum $C$ introduced  in eqn. (\ref{sup}) is finite. 
Then $C$ defines a squared norm $||\varphi||^2$ on $\Omega_0(\mathbb{H})$. The subspace  
$\Omega_0(\mathbb{H})$ is complete with respect to this norm. This norm is Hilbert, {\it i.e.}, 
it verifies the parallelogram identity. Hence the scalar product $\langle\varphi|\psi\rangle$  
defined on $\Omega_0(\mathbb{H})$ through
\begin{equation} 
4\langle \varphi|\psi\rangle:=
||\psi+\varphi||^2-||\psi-\varphi||^2+{\rm i}\,||\psi+{\rm
i}\,\varphi||^2-{\rm i}\,||\psi-{\rm i}\,\varphi||^2
\label{polar}
\end{equation}
turns the complete normed space $\Omega_0(\mathbb{H})$ into a Hilbert space with respect to the
scalar product (\ref{polar}). In fact,  via the HFT, the subspace $\Omega_0(\mathbb{H})$ is
isometrically isomorphic to the Hilbert space $L^2(0,\infty)$.

\section{Quantum operators from the HFT}\label{qmhft}

Introducing Planck's constant $\hbar$, the HFT reads
\begin{eqnarray}
\varphi(z)&=&{1\over \sqrt{2\pi\hbar}}\int_0^{\infty}{\rm d} s\, f(s)\,{\rm e} ^{{{\rm i}\over
\hbar}sz}\cr 
f(s)&=&{1\over \sqrt{2\pi\hbar}}\int_{-\infty}^{\infty}{\rm d} z\, \varphi(z)\, 
{\rm e}^{-{ {\rm i}\over \hbar}sz}.
\label{qreal}
\end{eqnarray}
In this section we promote the variables $z\in\mathbb{H}$ and $s\in (0, \infty)$ to quantum operators $Z$ and $S$, respectively, and study their properties.

We define operators $S$ and $Z$ as 
\begin{equation}
(Sf)(s):=s\,f(s), \qquad (Zf)(s):={\rm i}\hbar\,{{\rm d} f\over {\rm d} s}.
\label{pdef}
\end{equation}
Equation (\ref{qreal}) implies that a conjugate representation for them is given by their HFT transform,
\begin{equation}
(S\varphi)(z)=-{\rm i}\hbar\, {{\rm d}\varphi\over {\rm d} z},\qquad (Z\varphi)(z)=z\,\varphi(z).
\label{zdef}
\end{equation}
Irrespective of the representation chosen we have that the Heisenberg algebra 
\begin{equation}
[Z, S]= {\rm i}\hbar\,{\bf 1}
\label{hei}
\end{equation}
holds on the intersection $D(S)\cap D(Z)$ of their respective domains. Next we make precise what these domains are.

On the domain
\begin{equation} 
D(S)=\{f\in L^2(0,\infty):\; \int _0^{\infty}{\rm d} s\, s^2 
|f(s)|^2<\infty\},
\label{domq}
\end{equation}
which is dense in $L^2(0,\infty)$, the operator $S$ is symmetric,
\begin{equation}
\langle f|S|g\rangle^*=\langle g|S|f\rangle.
\label{barp}
\end{equation} 
A closed, symmetric, densely defined operator admits a selfadjoint extension if and only if 
its defect indices $d_{\pm}$ are equal. Moreover, such an operator is essentially selfadjoint
if and only if its defect indices are both zero \cite{YOSIDA}.  The operator $S$ 
turns out to be essentially selfadjoint, with  point, residual 
and continuous spectra given by
\begin{equation}
\sigma_p(S)=\phi, \qquad \sigma_r(S)=\phi, \qquad \sigma_c(S)=[0,\infty).
\label{pspectra}
\end{equation}

The properties of the conjugate operator $Z$ are subtler. One finds 
\begin{equation} 
{\langle f|Z| g\rangle }^*={\rm i}\hbar\, f(0)g^*(0) +
\langle g|Z| f\rangle,
\label{barzop}
\end{equation} 
so $Z$ is symmetric on the domain 
\begin{equation}
D(Z)=\{ f\in L^2(0,\infty): 
f\; {\rm abs.}\; {\rm cont.}, 
\int_0^{\infty}{\rm d} s\,|{{\rm d} f\over {\rm d} s}|^2 <\infty,\; 
f(0)=0\}.
\label{domz}
\end{equation}
($f$ is absolutely continuous). 
The adjoint $Z^{\dagger}$ also acts as ${\rm i}\,\hbar\,{\rm d}/{\rm d} s$, with a
domain $D(Z^{\dagger})$
\begin{equation} 
D(Z^{\dagger})=\{ f\in L^2(0,\infty): f\; {\rm abs.}\; {\rm cont.}, 
\int_0^{\infty}{\rm d} s\,|{{\rm d} f\over {\rm d} s}|^2 
<\infty\},
\label{domzdag}
\end{equation}
where the boundary condition $f(0)=0$ has been lifted. On the space $L^2(0,\infty)$ 
we have $d_{+}(Z)=0$, $d_{-}(Z)=1$. We conclude that  $Z$ admits no selfadjoint extension. 
Its point, residual and continuous spectra are
\begin{equation}
\sigma_p(Z)=\phi, \qquad \sigma_r(Z)=\mathbb{H}\cup \mathbb{R}, \qquad 
\sigma_c(Z)=\phi.
\label{zspectra}
\end{equation}

The domain $D(Z)$ is strictly contained in $D(Z^{\dagger})$. This implies that the operators
$X:=(Z+Z^{\dagger})/2$ and $Y:=(Z-Z^{\dagger})/2{\rm i}$ which one would naively construct
out of $Z$  are ill defined.  There is no way to define selfadjoint operators $X$ and $Y$ 
corresponding to the classical coordinates $x={\rm Re}\,z$ and $y={\rm Im}\, z$. This is compatible with the fact that,
the defect indices of $Z$ being unequal, $Z$ does not commute with any complex conjugation on
$L^2(0,\infty)$ \cite{YOSIDA}. However, we will see presently that one can 
make perfectly good sense of a quantum--mechanical operator $Z$ 
admitting no selfadjoint extension. 

With our choice of domain $D(Z)$, which makes $Z$ symmetric, $Z^2$ is also symmetric.
One proves that $d_-(Z^2)=1=d_+(Z^2)$. Hence $Z^2$, although not essentially selfadjoint, 
admits a selfadjoint extension. A popular choice is the Friedrichs extension \cite{YOSIDA}. 
Given an operator $A$, this extension is characterised by a boundedness condition
\begin{equation}
\langle \psi|A|\psi\rangle \geq -\alpha \,||\psi||^2 \qquad \forall \psi\in D(A)
\label{friedbound}
\end{equation}
for a certain $\alpha\geq 0$. Now the operator $Z^2$ admits a Friedrichs extension $Z^2_F$ 
with a lower bound
$\alpha=0$:
\begin{equation}
\langle f|Z^2_F|f\rangle \geq 0, \qquad \forall f\in D(Z^2_F).
\label{pbound}
\end{equation}
The point, residual and continuous spectra of this extension are
\begin{equation}
\sigma_p(Z^2_F)=\phi,\qquad \sigma_r(Z^2_F)=\phi, \qquad \sigma_c(Z^2_F)=[0,\infty).
\label{ptwospectra}
\end{equation}

Now the crucial point is that the square root of the Friedrichs extension 
allows us to define a selfadjoint momentum operator. 
Let us define the new operator $Z_{\sqrt{}}$ 
\begin{equation}  
Z_{\sqrt{}}:=+\sqrt{Z^2_F}.
\label{square}
\end{equation}  
$Z_{\sqrt{}}$ is selfadjoint, with a domain $D(Z_{\sqrt{}})$ univocally
determined by the spectral decomposition of $Z$ \cite{YOSIDA}. 
The point, residual and continuous spectra of $Z_{\sqrt{}}$ are
\begin{equation}
\sigma_p(Z_{\sqrt{}})=\phi,\qquad \sigma_r(Z_{\sqrt{}})=\phi,\qquad
\sigma_c(Z_{\sqrt{}})=[0,\infty).
\label{spectrappm}
\end{equation}
We observe that the operation of taking the Friedrichs extension does not commute with the operation of taking the square root. 

{}Finally let us consider transforming the operators $S$ and $Z$ under $SL(2,\mathbb{R})$.
We can reparametrise the coordinate $z\in \mathbb{H}$ by means of a M\"obius transformation
$z\mapsto \tilde z=(az+b)(cz+d)^{-1}$, with $ad-bc=1$.  Then $\tilde z\in\mathbb{H}$. We now write the HFT as
\begin{eqnarray}
\tilde\varphi(\tilde z)&=&{1\over \sqrt{2\pi\hbar}}
\int_0^{\infty}{\rm d} \tilde s\, \tilde f(\tilde s)\,
{\rm e}^{{{\rm i}\over\hbar}\tilde s\tilde z}\cr 
\tilde f(\tilde s)&=&{1\over \sqrt{2\pi\hbar}}
\int_{-\infty}^{\infty}{\rm d} \tilde z\,\tilde\varphi(\tilde z)\,{\rm e}^{-{ {\rm i}\over\hbar}
\tilde s\tilde z},
\label{tildeqreal}
\end{eqnarray}
where $\tilde s\in (0,\infty)$ is the variable conjugate to $\tilde z$ under (\ref{tildeqreal}). 
One can define quantum operators $\tilde S$ and $\tilde Z$ satisfying 
the Heisenberg algebra (\ref{hei}). Hence this is a canonical transformation from $S,Z$ 
to $\tilde S, \tilde Z$. In terms of the transformed variables $\tilde s$, $\tilde z$, the transformed operators $\tilde S$ and $\tilde Z$ 
have the same spectra as before.

\section{Examples and discussion}\label{discussion}

The standard Fourier transformation maps (a subspace of) $L^2(\mathbb{R})$ into (a subspace of) $L^2(\mathbb{R})$. It is also an isospectral transformation between selfadjoint operators such as the position operator $X$ and its conjugate momentum operator $P$ for a particle on the whole real line $\mathbb{R}$. In the context of the standard Fourier transformation on $L^2(\mathbb{R})$, coordinate and momentum are sometimes referred to as a {\it Schr\"odinger pair}. On the contrary, the HFT is not an isospectral transformation: the operators $S$ and $Z$ do not have identical spectra. Furthermore, the very choice of the dynamical variable to be represented by complex variable $z$ of the HFT is a nontrivial choice in itself. 

Since the Hamiltonian $H$ is bounded from below it makes sense to take, in section \ref{qmhft}, the selfadjoint operator $S$ as the Hamiltonian $H$ and the nonhermitian operator $Z$ as the time operator $T$. In this way  we arrive at a selfadjoint time operator $T_{\sqrt{}}:=\sqrt{T_F^2}$ with the semiaxis $(0, \infty)$ as its continuous spectrum. It is this latter operator $T_{\sqrt{}}$ that we take to define (positive) time. We further observe that we have an additional $SL(2, \mathbb{R})$ symmetry at our disposal, generated by translations, dilations and inversions acting on $\mathbb{H}$ and hence also on its boundary $\mathbb{R}$. Under dilations $x\mapsto \lambda x$, where $\lambda > 0$, the semiaxis $(0, \infty)$ transforms into itself, while we can shift it into any desired interval $(k, \infty)$, $k\in\mathbb{R}$, by means of a translation $x\mapsto x-k$. Under an inversion $x\mapsto -1/x$, the semiaxis $(0, \infty)$ transforms into its opposite $(-\infty, 0)$.  Convening that the inversion maps the point at infinity into zero, and viceversa, it suffices to consider the inversion transformation and its corresponding operator $\tilde T$ in order to obtain the whole real line $\mathbb{R}$ as the (joint) continuous spectrum of the two time operators $T_{\sqrt{}}$ and $\tilde T_{\sqrt{}}$.
Overall there is a whole $SL(2, \mathbb{R})$'s worth of time operators to choose from. This fits in well with the multiplicity of existing time--of--arrival operators in the literature \cite{TIME, EGUSQUIZA, GALAPON}, although a general criterion to map a given $SL(2, \mathbb{R})$--time, as proposed here, with those of refs. \cite{TIME, EGUSQUIZA, GALAPON}, is lacking. 

The existenc of a whole $SL(2, \mathbb{R})$'s worth of time operators brings us to a related question, namely, whether or not our formalism also works in the presence of degeneracy. The answer is affirmative. $SL(2, \mathbb{R})$ acts as per eqn. (\ref{tildeqreal}). This group has finite--dimensional representations in all real dimensions, as well as a continuous series of infinite--dimensional representations. Its action commutes with the operator $S$ in the sense that the transformed operator $\tilde S$ has the same spectrum as $S$, even if the corresponding eigenfunctions (eqns. (\ref{ramalloquetefollen}), (\ref{ramallokomprateunchampuantikaspa}) below) get exchanged under an $SL(2, \mathbb{R})$--transformation. Thus picking a representation of $SL(2, \mathbb{R})$ with the desired dimension ({\it i.e.}, with the desired degeneracy), eventually infinite, the commutativity of the $SL(2, \mathbb{R})$--action with the Hamiltonian $H=S$ ensures that our formalism remains valid also in the presence of degeneracy.

As an example let us work out the case of a free particle moving on the whole real line. Standard quantum mechanics tells us that the Hamiltonian $H_f=P^2/2m$ is twofold degenerate. In coordinate representation, the eigenfunctions corresponding to the eigenvalue $E_p=p^2/2m$ are $u_{\pm    p}(x)={\rm exp}\left(\pm{\rm i}px\right)$. This twofold degeneracy can be understood in terms of $SL(2, \mathbb{R})$--transformations as follows.
The spectrum of $H_f$ is the semiaxis $[0,\infty)$, so one naturally identifies $H_f$ with the operator $S$ of eqn. (\ref{pdef}). Now let us use the HFT and set $S=H_f$ in eqn. (\ref{pdef}). As usual we denote by $E, t$ the variables corresponding to the operators $H_f, T_{\sqrt{}}$, respectively. In energy representation the eigenfunction of $Z=T_{\sqrt{}}$ corresponding to the eigenvalue $t'$ is 
\begin{equation}
f_{t'}(E)={\rm exp}\left(-\frac{{\rm i}}{\hbar}Et'\right),
\label{ramalloketemetanunpaloporelkulo}
\end{equation}
while that of $H_f$ corresponding to the eigenvalue $E'$ is
\begin{equation}
f_{E'}(E)=\delta(E-E'). 
\label{ramalloquetefollen}
\end{equation}
We can HFT--transform the eigenfunction (\ref{ramalloquetefollen}) to obtain its expression in time representation, 
\begin{equation}
\varphi_{E'}(t)={\rm exp}\left(\frac{{\rm i}}{\hbar}E't\right).
\label{ramallokomprateunchampuantikaspa}
\end{equation}
The twofold degeneracy of the eigenfunctions (\ref{ramallokomprateunchampuantikaspa}) is recovered once one remembers that the operator $T_{\sqrt{}}$ is defined as the {\it positive}\/ square root (\ref{square}); negative times are obtained from the $SL(2,\mathbb{R})$--transformed operator $\tilde T_{\sqrt{}}$ discussed previously. We further notice that the eigenvalue equation satisfied by the eigenfunctions (\ref{ramallokomprateunchampuantikaspa}), 
\begin{equation}
H_f\varphi_{E'}(t)=-{\rm i}\hbar\frac{{\rm d}\varphi_{E'}}{{\rm d}t}=E'\varphi_{E'}(t),
\label{marikonbarbon}
\end{equation}
is actually equivalent to the time--dependent Schr\"odinger equation, since the minus sign multiplying ${\rm i}\hbar$ can be flipped by changing to the $SL(2,\mathbb{R})$--transformed time $\tilde T_{\sqrt{}}$. This is in agreement with the fact that the time--dependent Schr\"odinger equation can be understood as the operator identity
\begin{equation}
{\rm i}\hbar\frac{\partial}{\partial t}=-\frac{\hbar^2}{2m}\nabla^2+V=H.
\label{psoedemierda}
\end{equation}

Beyond the particular case when $H=H_f$, however, the identification $H=S$ must be made with some care. Usually the basic operators, {\it i.e.}, those satisfying the Heisenberg algebra, are position $X$ and momentum $P$, and the Hamiltonian $H$ is a function of the latter, typically $H=P^2/2m+V(X)$. As $X$ and $P$ do not commute, the time--independent Schr\"odinger equation $Hu(x)=Eu(x)$ is nontrivial, but there are two cases when diagonalising the Hamiltonian reduces to diagonalising one of the two basic operators $X,P$, as in eqns. (\ref{ramalloketemetanunpaloporelkulo})--(\ref{ramalloquetefollen}). The first case is when the kinetic term $P^2/2m$ can be neglected in comparison with the potential term $V(X)$; then it suffices to diagonalise $X$. The second case is when $V=0$ identically, so we need only diagonalise $P$: this is the free particle. As a rule, however, our particle will not be free, and the spectrum of $H$ will only share one property in common with the spectrum of $S$, namely, that they are both bounded from below. While a constant shift in energy (an $SL(2,\mathbb{R})$--transformation) can bring the ground state of $H$ to zero energy, in general the spectrum of $S$ will be larger than that of $H$. All this notwithstanding, the spectral objections just raised to the identification $S=H$ can also arise (and actually {\it do}\/ arise) for the usual Schr\"odinger pair $X,P$ in the presence of a potential $V$, or in the presence of boundary conditions. Thus, {\it e.g.}, while $P=-{\rm i}\,\hbar\,{\rm d}/{\rm d}x$ on its own has all of $\mathbb{R}$ as its spectrum, this is generally no longer true when $V\neq 0$ ({\it e.g.}, a particle inside an infinite potential well). In other words, the previous {\it caveat}\/ concerning the spectra must also be borne in mind when approaching quantum mechanics from the point of view based on the usual Fourier transformation and the Heisenberg algebra $[X,P]={\rm i}\hbar$.

To summarise, the HFT allows one to construct a selfadjoint Hamiltonian bounded from below that is conjugate to a time operator whose spectrum is the whole real line. As opposed to the usual Fourier transformation, which is an isospectral transformation between $X$ and $P$, the HFT is not an isospectrality between $H$ and $T$. In compensation, there is an $SL(2,\mathbb{R})$--symmetry acting on $T$ which the usual Fourier transformation lacks. Our formalism is based on the Heisenberg algebra $[T, H]={\rm i}\hbar$ rather than the usual $[X,P]={\rm i}\hbar$. We have proved the equivalence between the  approach based on the HFT with the Heisenberg algebra $[T, H]={\rm i}\hbar$ and the approach based on the usual Fourier transformation with the Heisenberg algebra $[X,P]={\rm i}\hbar$.

{\bf Acknowledgements}

It is a great pleasure to thank J. de Azc\'arraga for encouragement and support, I. Egusquiza for technical discussions, and the referees for useful suggestions to improve the manuscript. The author thanks  Max-Planck-Institut f\"ur Gravitationsphysik, Albert-Einstein-Institut (Potsdam, Germany), where this work was was begun, for hospitality. This work has been partially supported by European Union network MRTN--CT--2004--005104, by research grant BFM2002--03681 from Ministerio de Ciencia y Tecnolog\'{\i}a, by research grant GV2004--B--226 from Generalitat Valenciana, by EU FEDER funds, by Fundaci\'on Marina Bueno and by Deutsche Forschungsgemeinschaft.

\end{document}